\documentclass[aps,prb,twocolumn,superscriptaddress,showpacs,epsf]{revtex4-1}

\usepackage{rotating}
\usepackage{graphicx}
\usepackage{latexsym}
\usepackage{amssymb}
\usepackage{amsfonts}
\usepackage{soul}
\usepackage{color}

\begin{document}
\mathchardef\mhyphen="2D
% Use the \preprint command to place your local institutional report
% number in the upper righthand corner of the title page in preprint mode.
% Multiple \preprint commands are allowed.
% Use the 'preprintnumbers' class option to override journal defaults
% to display numbers if necessary
%\preprint{}

%Title of paper
\title{A closer look at the low frequency dynamics of vortex matter}

\author{B. Raes}
\affiliation{INPAC -- Institute for Nanoscale Physics and Chemistry, Department of Physics and Astronomy , KU Leuven, Celestijnenlaan 200D, B--3001 Leuven, Belgium}
\affiliation{Current address -- Physics and engineering of nanodevices group, Catalan institute of Nanoscience and Nanotechnology (ICN), Campus UAB, E-08193
Bellaterra, Spain.}

\author{C. de Souza Silva}
\affiliation{Departamento de Fisica, Universidade Federal de Pernambuco, Cidade Universitaria, 50670-901 Recife-PE, Brazil}

\author{A.V. Silhanek}
\affiliation{D\'{e}partement de Physique, Universit\'{e} de Li\`{e}ge, All\'{e}e du 6 ao\^{u}t 17, B5, B--4000 Sart Tilman, Belgium}

\author{L.R.E. Cabral}
\affiliation{Departamento de Fisica, Universidade Federal de Pernambuco, Cidade Universitaria, 50670-901 Recife-PE, Brazil}

\author{V.V. Moshchalkov}
\affiliation{INPAC -- Institute for Nanoscale Physics and Chemistry, Department of Physics and Astronomy , KU Leuven, Celestijnenlaan 200D, B--3001 Leuven, Belgium}

\author{J. Van de Vondel}
\affiliation{INPAC -- Institute for Nanoscale Physics and Chemistry, Department of Physics and Astronomy , KU Leuven, Celestijnenlaan 200D, B--3001 Leuven, Belgium}

\date{\today}

\begin{abstract}

Using scanning susceptibility microscopy, we shed new light on the dynamics of individual superconducting vortices and examine the hypotheses of the phenomenological models traditionally used to explain the macroscopic ac electromagnetic properties of superconductors. The measurements, carried out on a 2H-NbSe$_2$ single crystal at relatively high temperature $T=6.8$ K, show a linear amplitude dependence of the global ac-susceptibility for excitation amplitudes between 0.3 and 2.6 Oe. We observe that the low amplitude behavior, typically attributed to the shaking of vortices in a potential well defined by a single, relaxing, Labusch constant, corresponds actually to strongly non-uniform vortex shaking. This is particularly accentuated  in the field-cooled disordered phase, which undergoes a dynamic reorganization above 0.8 Oe as evidenced by the healing of lattice defects and a more uniform oscillation of vortices. These observations are corroborated by molecular dynamics simulations when choosing the microscopic input parameters from the experiments. The theoretical simulations allow us to reconstruct the vortex trajectories  providing deeper insight in the thermally induced hopping dynamics and the vortex lattice reordering.
\end{abstract}

% insert suggested PACS numbers in braces on next line

\maketitle

\section{Introduction}
\label{sec.introduction}

The low frequency response of type-II superconductors to electromagnetic excitations is ruled by the dynamics of quantum units of magnetic flux, so called vortices~\cite{Pompeo2008}. These are three dimensional elastic entities interacting repulsively, typically  immersed in a random environment of pinning centers. Moreover, in most cases the influence of thermal excitations cannot be neglected, especially in the technologically relevant high temperature superconductors, adding an extra ingredient to this already complex problem~\cite{Blatter1994}.

The competing vortex-vortex and vortex-pinning center  interactions can give rise to a vortex lattice poisoned with defects where the symmetry of the lattice is violated (e.g. disclinations). The healing of these defects can, under certain conditions, be obtained by submitting the vortex lattice to an external excitation. Indeed, it has already been shown both theoretically and experimentally, that a disordered vortex lattice resulting from a relatively strong random pinning distribution, can undergo a dynamical reordering transition when driven by a dc external force $F > F_{dp}$, where $F_{dp}$ is the depining force ~\cite{Koshelev1994, Yaron1995, Duarte1996, Ryu1996, Fangohr2001, Reichhardt2009}.  This transition not always consists of a monotonous and progressive healing of topological defects as the drive increases, but in some cases a maximum of disclinations in the vortex lattice is observed at the onset of depinning $F \sim F_{dp}$ \cite{Koshelev1994, Duarte1996}.

Although particular effort has been devoted to understand the dynamic behavior under dc drive, somewhat less attention has been paid to ac excitations~\cite{Henderson1998, Paltiel2000, Valenzuela2001, Valenzuela2002, Pasquini2008, daroca2010, Mangan2008}. Unfortunately, the extrapolation of the findings obtained under dc drive to predict the ac dynamics is not always straightforward. For instance, it has been reported that for similar excitation amplitude, dc experiments can induce disorder in the vortex lattice while ac shaking leads to ordering~\cite{Henderson1998, Paltiel2000, Valenzuela2001, Valenzuela2002}.

Despite the continuous progress made during the last decades, our current understanding of the complex dynamic behavior of vortex lattices relies on observables involving a statistical average over a large number of vortices\cite{Gomory1997} or, at best, through local static imaging\cite{Duarte1996, Fasano2002, Menghini2002, Marchevsky1998}. Global measurements rely on introducing certain assumptions on the {\it average} vortex motion thus losing the details of {\it individuals}, very much like bridging thermodynamics to statistical physics. For instance, the surface impedance of the superconducting material at low ac amplitudes can be deduced from the assumption that the coupling between vortices and pinning centers can be modeled by a single and isotropic spring constant known as the Labusch constant ~\cite{Coffey1991,Brandt1991,vanderbeek1993}. The expected macroscopic response is then determined by combing the obtained complex impedance with both Faraday's and Amp\`{e}re's law.

The question now arises as to whether the simplified hypotheses used in these models are actually valid at the microscopic level. It is difficult to find the answer to this question based on static imaging, since snapshots lack the time variable, essential to track the history encoded in the vortex trajectories and to unveil the characteristic time scales involved in the vortex hopping.

%Figure 0
\begin{figure*}[t]
\hspace{2.3cm}\includegraphics[width=0.9\linewidth]{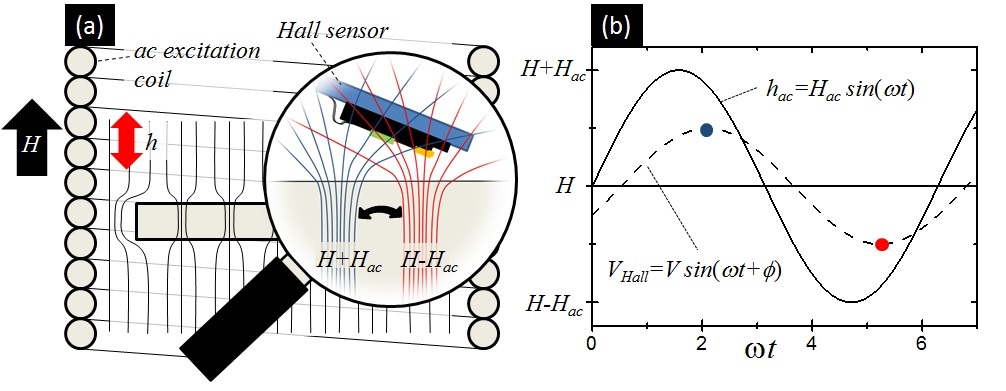} \caption{ Figure 1. (Color online) (a) Schematic overview of the scanning susceptibility microscopy setup. A superconducting sample is placed in a dc magnetic field, $H$, generated by a superconducting coil surrounding a collinear copper coil generating an ac field $h_{ac}(t)$.  The time averaged magnetic field profile due to the present vortices and the screening currents is schematically shown by the black lines.  The magnifying glass provides a closer look to the induced ac-vortex motion. When the drive is small, the ac magnetic field induces a periodic force on the vortices shaking them back and forth. A Hall sensor picks up locally the associated time dependent Hall voltage, $V_{Hall}$. A lock-in amplifier, provided  with both $h_{ac}(t)$ as a reference and $V_{Hall}$
 are shown in (b),  allows one to extract both the in-phase, $b_1'(x,y)$,  and the out-of phase, $b_1''(x,y)$,  components of the local magnetic response.
 }
\label{fig:setup}
\end{figure*}

In this work we investigate the ac dynamics of individual superconducting vortices by going beyond static imaging. We have chosen to perform these experiments on  NbSe$_2$, arguably one of the most extensively studied type-II superconductor. We monitored the average vortex distribution with scanning Hall probe microscopy (SHPM)\cite{Kirtley} and the local ac vortex dynamics using scanning ac-susceptibility microscopy (SSM)\cite{DeFeo,Kramer,Kramer1, Raes2012}. In particular, we use the combination of both techniques to map the development and evolution of the different dynamical states as a function of driving amplitude when starting from an initially disordered vortex lattice. The local observation and characterization of these dynamical states unveils a far more richer and complex scenario than the one pictured from the mean ac response using the aforementioned models. More precisely, a coarsening of topological defects initially present in a prepared disordered vortex state is observed {\it in situ} upon increasing the external ac magnetic field. This dynamical re-organization is strongly influenced by the thermal hopping dynamics of vortices, resulting in a much faster VL reordering as compared to the $T=0$ K case. In addition, in the disordered state the results reveal a highly non-uniform oscillatory motion reflecting the local anisotropic properties of the potential landscape felt by individual vortices whereas the ordered state exhibits a more coherent motion. The strong out-of-phase component of the vortex motion can be unambiguously attributed to the dissipative character of the thermally activated motion over the pinning potential. These experimental findings are corroborated by molecular dynamics simulation which shows an excellent agreement with the experimental results.

\section{Experimental details}

The sample under investigation is a 2H-NbSe$_2$ single crystal of approximate dimensions $2.5\times2.5\times 0.5$ mm$^3$, grown by a standard iodine vapor transport method\cite{Li}, which has a critical temperature of $T_c=7.05$ K  at zero magnetic field.

The vortex distribution is probed by measuring the $z$-component of the time averaged local induction, $\langle b_z(x,y,t)\rangle _t$, with a modified low-temperature scanning Hall probe microscope from Nanomagnetics Instruments. The typical scan area at 4.2 K is 16$\times$16 $\mu$m$^2$. The images were recorded in lift-off mode with the Hall sensor at about 1.5 $\mu$m above the surface of the sample \cite{Brisbois}. Additional $xy$-positioners allow us to explore different regions of the sample.

% Figure 1
\begin{figure*}[t]
\hspace{2.3cm}\includegraphics[width=0.85\linewidth]{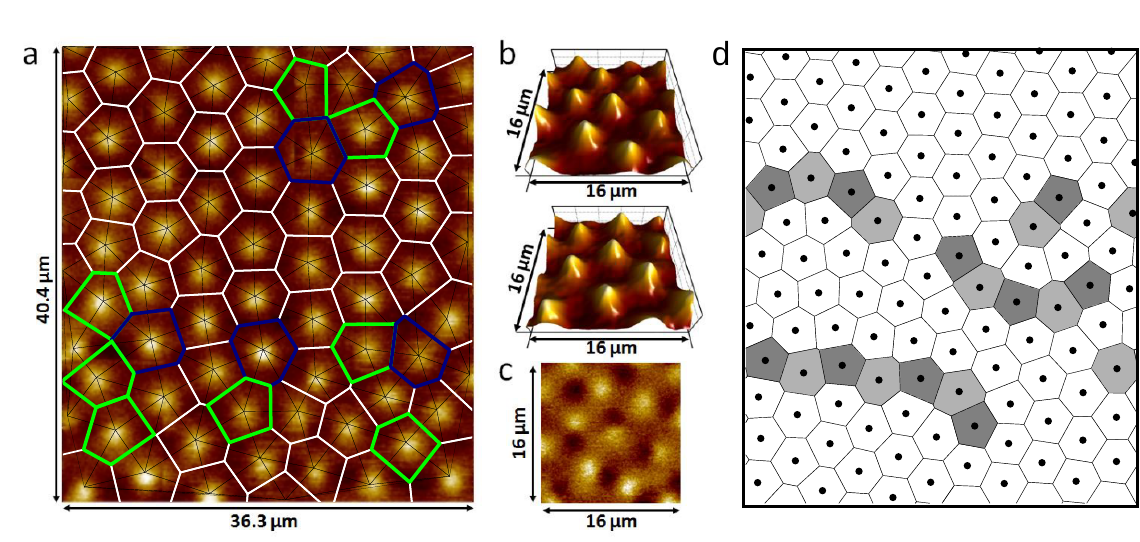} \caption{(Color online) (a) Vortex configuration at 4.2 K after a field-cooling experiment in 1 Oe applied perpendicular to the sample surface. The Voronoi construction is plotted on top of the image. (b) A sequence of two different field-cooling experiments in 0.9 Oe, imaged at 4.2K. (c) Difference image between the two aforementioned experiments at 0.9 Oe. (d) Typical vortex configuration obtained from molecular dynamics simulations on a weakly disordered 2D superconducting system (cf. section~\ref{sec.MD}).
 }
\label{fig:VLconf}
\end{figure*}

To probe the ac vortex dynamics, we continuously excite the sample with an external oscillating magnetic field, $h_{ac}(t)=H_{ac}\sin(\omega t)$, while picking up the Hall voltage induced by the time-varying local induction $b_z(x,y,t)$ as shown in Fig.\ref{fig:setup}. The collinear dc and ac external magnetic fields are always applied perpendicularly to the sample surface. The time dependent local induction, $b_z(x,y,t)\sim V_{Hall}(x,y,t)$ picked up by the Hall probe, is Fourier analyzed with a lock-in amplifier using the applied ac magnetic field as reference,
\begin{eqnarray}\label{Eq0}
&b_z(x,y,t)=\nonumber\\
&\sum_{n=1}^{\infty}[b_{n}'(x,y)  \sin(n \omega t) +b_{n}''(x,y)  \cos (n\omega t)]
\end{eqnarray}
The first term ($n=1$)  of the Fourier series in Eq.\ref{Eq0}, i.e. the in-phase, $b_{1}'(x,y)$, and out-of phase, $b_{1}''(x,y)$, Fourier components, are normally dominant and represent the linear response to the local variation of the magnetic induction, thus $b_z(x,y,t) \approx b_{1}'(x,y)  \sin( \omega t) +b_{1}''(x,y)  \cos(\omega t)\nonumber$. To avoid unwanted effects such as eddy current heating, the skin effect of the sample holder or the frequency dependence of the Hall probe sensitivity, we perform all measurements at a fixed low driving frequency of $f=$$\omega/2\pi$$=$77.123 Hz. The dwell time at every pixel ($\tau_{pix}$) and the integration time of the lock-in ($\tau_{int}$) are chosen appropriately ($\tau_{pix},\tau_{int}\gg1/f$) while the measured phase between the picked up signal and the ac magnetic field drive is set to zero above $T_c$.

\section{Experimental results}

\subsection{Generation of frozen weakly disordered vortex states}

An initial disordered vortex state is prepared following a field-cooling (FC) procedure, in which the sample temperature is decreased from above $T_c$ at a constant field, $H$, applied along the $c$ axis, down to 4.2 K. This final temperature is well below the so-called quenching temperature, $T_q$, at which the bulk pinning freezes the vortex lattice in a stable configuration~\cite{Marchevsky}. The dc fields in our experiments, $H\leq1$ Oe, are well below the field range where the peak-effect anomaly is observed in high-purity NbSe$_2$ samples~\cite{Menghini2002}.

Figure \ref{fig:VLconf}(a) shows a typical vortex configuration for $H=1.0$ Oe at $T=4.2$ K. The average flux density associated with such distribution is $B\simeq0.9$ G, which indicates a rather uniform flux distribution over the sample.  The vortex configuration corresponds to a weakly disordered distribution as consequence of the random pinning. As evidenced by the Voronoi diagram plotted on top of the image, within the observed area, most of the topological defects are bound pairs of positive (seven-fold defects highlighted in blue) and negative (five-fold defects in green) disclinations. Moreover, most of them are clustered together. Such morphology is consistent with previous imaging experiments on pure NbSe$_2$ single crystals~\cite{Marchevsky1998,Fasano2002} as well as with numerical simulations on 2D vortex systems interacting with random weak pinning distributions~\cite{Moretti2004,Chandran2004}. Figure~\ref{fig:VLconf}(b) shows the vortex configurations in the same sample region for two different runs of a FC procedure performed at $H=0.9$ Oe. When subtracting the two images in panel (b) the image shown in figure~\ref{fig:VLconf}(c) is obtained. We clearly see from this differential image that vortices occupy different positions for two independent FC runs. This suggests that the quenched disorder in the sample comprises a highly dense distribution of weak pinning centers, thus providing a multitude of energetically quasi-equivalent metastable states rather than favoring a particular configuration. For comparison, we show in figure \ref{fig:VLconf}(d) a typical vortex distribution obtained from the molecular dynamics (MD) simulations described in Section~\ref{sec.MD}. The root-mean-square strength of the random pinning potential was suitably tuned in order to have a frozen vortex state with a morphology similar to the experiment.

\begin{figure*}[t]
\hspace{2.54cm}
\includegraphics[width=0.65\linewidth]{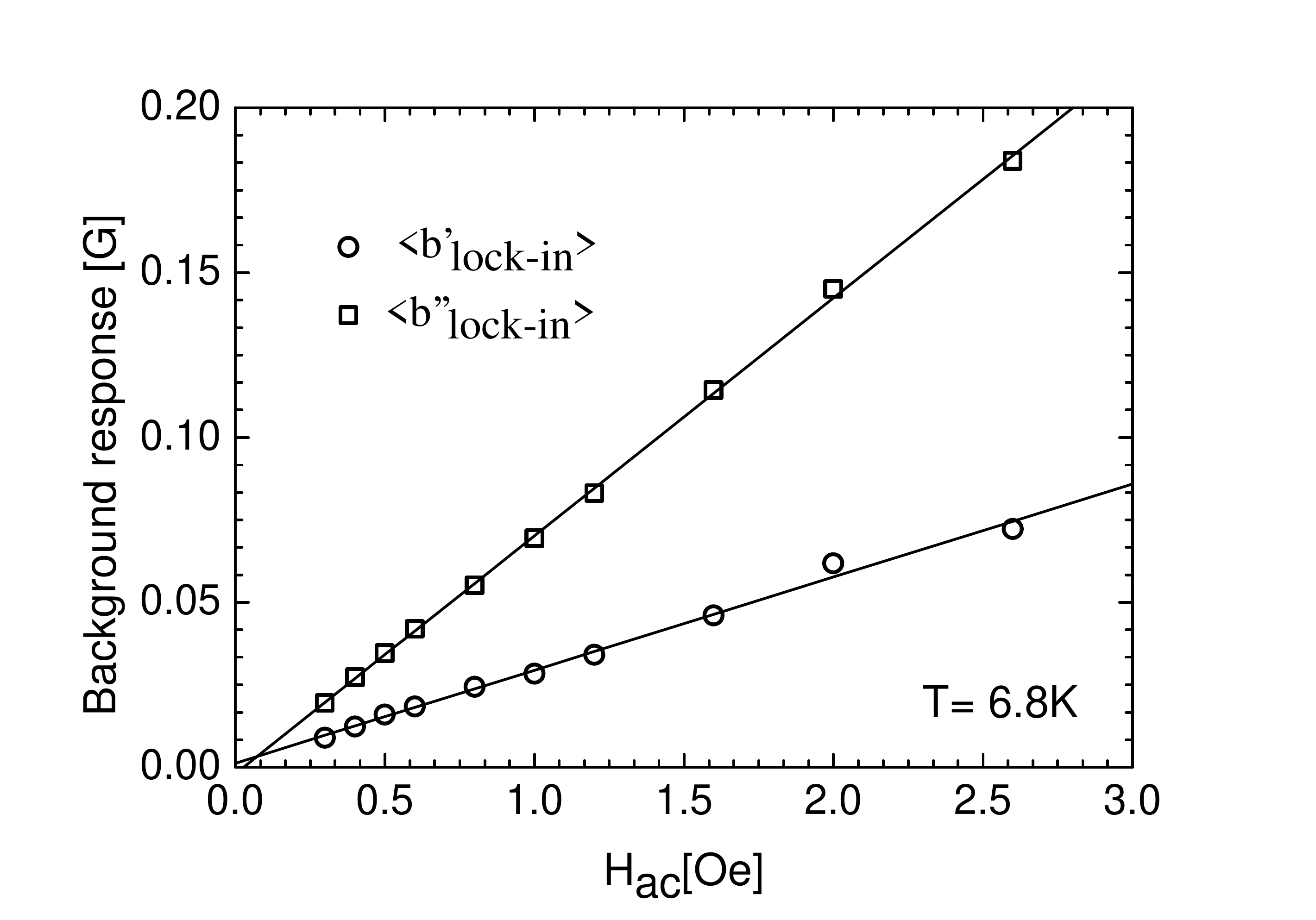}
\vspace{-5mm} \caption{In-phase (circles) and out-of-phase (squares) components of the background signal averaged over the scan area as functions of the ac amplitude $H_{ac}$ at $T=6.8$ K and after a $H=1.0$ Oe FC procedure. The  lines are linear fits to the data. } \label{fig.ExpMean}
\end{figure*}

\subsection{Mean ac response: identification of the dynamical regime}

Let us now identify the dynamical regime of the vortex state within the context of the macroscopic response theory, after a $H=1.0$ Oe FC procedure to $T=6.8$ K.  To that end, we investigate the amplitude dependence of the signal measured via the SSM technique {\it averaged over the scan area} for ac amplitude values ranging from 0.3 Oe to 2.6 Oe while keeping the dc field at 1.0 Oe. Figure~\ref{fig.ExpMean} shows the amplitude dependence of both $\langle b_\mathrm{lock{\mhyphen}in}'\rangle$ and $\langle b_\mathrm{lock{\mhyphen}in}''\rangle$ as detected by the lock-in amplifier. A clear linear dependence is observed, suggesting that the mean response lies within a well-defined linear dynamical regime characterized by a single phase-lag $\phi=\tan^{-1}(\langle b_\mathrm{lock{\mhyphen}in}''\rangle/\langle b_\mathrm{lock{\mhyphen}in}'\rangle)$. Its value, as calculated from least square fits (solid lines in the figure), $\phi=(-66.7\pm1.4)^\circ$, is far from zero thus revealing a strongly dissipative dynamics. Notice that there is no hint in this average response of a possible dynamic transition in the vortex lattice.

In the literature on linear vortex response, one usually considers an averaged version of the equation of motion where the coupling among vortices and between vortices and pinning centers are represented by a single scalar, field-dependent spring constant, $\alpha_L$, called the Labusch constant~\cite{Labush1969}. In this model, vortex displacements $\mathbf{u}$ are assumed to be parallel to the applied drive and the restoring force is simply given by
 \begin{equation}\label{eq.restoring1}
     \mathbf{F}^{\rm res}=-\alpha_L\mathbf{u},
 \end{equation}
 which, ignoring thermal fluctuations, yields the vortex response in the frequency domain\cite{Gittleman1966}:
 \begin{eqnarray}
 \mathbf{u}(\omega) &=& \chi(\omega)\mathbf{F}_{ac} \label{eq.resp-mean}\\
 \chi(\omega) &=& \left(\alpha_L-i\omega\eta\right)^{-1}. \label{eq.respfunc-mean}
 \end{eqnarray}
Here the dispersive vortex response function $\chi(\omega)$ is a complex scalar and $\eta$ is the viscous drag coefficient induced by dissipative processes of quasiparticles within the vortex core. This induces a viscous drag force opposing vortex motion, which introduces a phase lag with respect to the drive given by $\tan^{-1}(\tau_p\omega)$, where $\tau_p=\eta/\alpha_L$ is the inverse pinning frequency.  This force stands out from the others only when vortices acquire high speed. This effect only becomes appreciable when either the driving frequency or the drive amplitude are high enough. For high frequencies (typically microwave) vortices tend to shake inside pinning centers making tiny displacements, in such a way that the restoring force can be neglected. In the limit of strong drive, vortices move past many pinning sites at a high speed and the pinning potential is washed out~\cite{Koshelev1994, Schmid1973, deSouzaSilva2002}. Since in our experiments both frequency and amplitude are small, a more plausible scenario for the observed dissipation is a linear dynamical regime dominated by thermally activated vortex hopping.

As pointed out by Brandt~\cite{Brandt1991}, thermally activated vortex hopping from one pinning site to another results in a relaxing Labusch parameter $\alpha_L(t)=\alpha_Le^{-t/\tau}$, where $\tau$ is the relaxation time determined by the Arrhenius form $\tau\sim \tau_pe^{U_0/kT}$ ($U_0$ represent the typical value of the pinning energy barriers). Within linear response theory, such time-dependent restoring force constant is accounted for by the complex  parameter $\alpha_L/(1-i/\omega\tau)$~\cite{Coffey1991,Brandt1991,vanderbeek1993}. The general solution for the linear vortex response including thermally assisted hopping is given by the real part of $\mathbf{u}(\omega, T)e^{i\omega t}$ with
 \begin{eqnarray}
 \mathbf{u}(\omega, T) &=& \chi(\omega, T)\mathbf{F}_{ac} \label{eq.resp-mean}\\
 \chi(\omega, T) &=& \left(\frac{\alpha_L}{1-i/\omega\tau} - i\eta\omega\right)^{-1}. \label{eq.respfunc-mean}
 \end{eqnarray}

 Indeed, within the mean-field Coffey-Clem-Brandt model  [Eqs. (\ref{eq.resp-mean}) and (\ref{eq.respfunc-mean})] and assuming that the excitation frequency lies in the regime $\omega\ll\alpha_p/\eta$, the mean magnetic permeability can be estimated as $\mu(\omega)=\langle b\rangle/h_{ac}\propto 1-i/\omega\tau$, from which we obtain the mean hoping time $\tau=1/[\omega\tan(-\phi)]=0.83\pm0.06$ ms. For the excitation fields used in our experiment one period of the external force spans about $15.5\tau$. This result is consistent with the previously assumed thermal hopping scenario~\cite{Raes2012}.

  However, one should keep in mind that such analysis represents a statistical average and, in general, it is not valid on the scale of single vortex dynamics. In fact, for a disordered vortex arrangement, the energy landscape probed locally by a vortex as a result of interactions with other vortices and with pinning centers is far from isotropic. Moreover the motion of each vortex couples to the motion of its neighbors as a result of the
 non-locality of vortex-vortex interactions. %Such coupling lead to
 %the excitation of collective modes, which can ultimately lead to
 %structural changes in the vortex configurations resulting in
 %nonlinear response.
Therefore, it is clear that \emph{a single coupling constant depending only on an average pinning force and the local flux density is insufficient to accurately describe the dynamics on a local scale}.

\subsection{Local ac response: evidence of dynamical ordering}

In principle, the signals $\langle b_\mathrm{lock{\mhyphen}in}'\rangle$ and $\langle b_\mathrm{lock{\mhyphen}in}''\rangle$ contain the response of the screening currents plus the average vortex response. In order to isolate the local ac response produced only by the vortices within a scan area, two steps are needed. Firstly, {\it for each probe position}, we subtract from the measured signal the background response generated by all currents except those encircling the vortices within the scan area. Notice that vortices moving back and forth without ever leaving the scan area have a negligible contribution to the net ac response and hence the background signal is approximately uniform and can be estimated as $b_{\rm bkg} \approx \langle b_\mathrm{lock{\mhyphen}in}'\rangle + i\langle b_\mathrm{lock{\mhyphen}in}''\rangle$. Secondly, it is convenient to refer to the dephasing of the signal with respect to the actual Lorentz force that drives the vortices in the scan area  instead of the applied field $h_{ac}(t)$. Such force can be estimated as $\mathbf{F}_L=\Phi_0\hat{\mathbf{z}}\times\mathbf{j}_{\rm bkg}$, where $\mathbf{j}_{\rm bkg}=\nabla\times\mathbf{b}_{\rm bkg}(\mathbf{r},t)$, and thus has the time dependence ${b}_{\rm bkg}\sim \cos(\omega t+\phi)$. Following this two-step procedure it is possible to obtain the in-phase, $b_v'$, and out-of-phase, $b_v''$, components representing the local response of the vortices by subtracting the background from the measured signal and rotating the result by $\phi$, that is:
%%
%\begin{eqnarray}
% b_v' &=& (b_\mathrm{lock{\mhyphen}in}'-\langle b_\mathrm{lock{\mhyphen}in}'\rangle)\cos\phi -  (b_\mathrm{lock{\mhyphen}in}''-\langle b_\mathrm{lock{\mhyphen}in}''\rangle)\sin\phi \nonumber\\
% b_v'' &=& (b_\mathrm{lock{\mhyphen}in}'-\langle b_\mathrm{lock{\mhyphen}in}'\rangle)\sin\phi +  (b_\mathrm{lock{\mhyphen}in}''-\langle b_\mathrm{lock{\mhyphen}in}''\rangle)\cos\phi
%\end{eqnarray}
\begin{eqnarray}
 b_v' = (b_\mathrm{lock{\mhyphen}in}'-\langle b_\mathrm{lock{\mhyphen}in}'\rangle)\cos\phi\nonumber\\
 -  (b_\mathrm{lock{\mhyphen}in}''-\langle b_\mathrm{lock{\mhyphen}in}''\rangle)\sin\phi \nonumber\\
 b_v''= (b_\mathrm{lock{\mhyphen}in}'-\langle b_\mathrm{lock{\mhyphen}in}'\rangle)\sin\phi\nonumber\\
  +  (b_\mathrm{lock{\mhyphen}in}''-\langle b_\mathrm{lock{\mhyphen}in}''\rangle)\cos\phi
\end{eqnarray}

\begin{figure*}[t]
%\hspace{2.54cm}
\includegraphics[width=\linewidth]{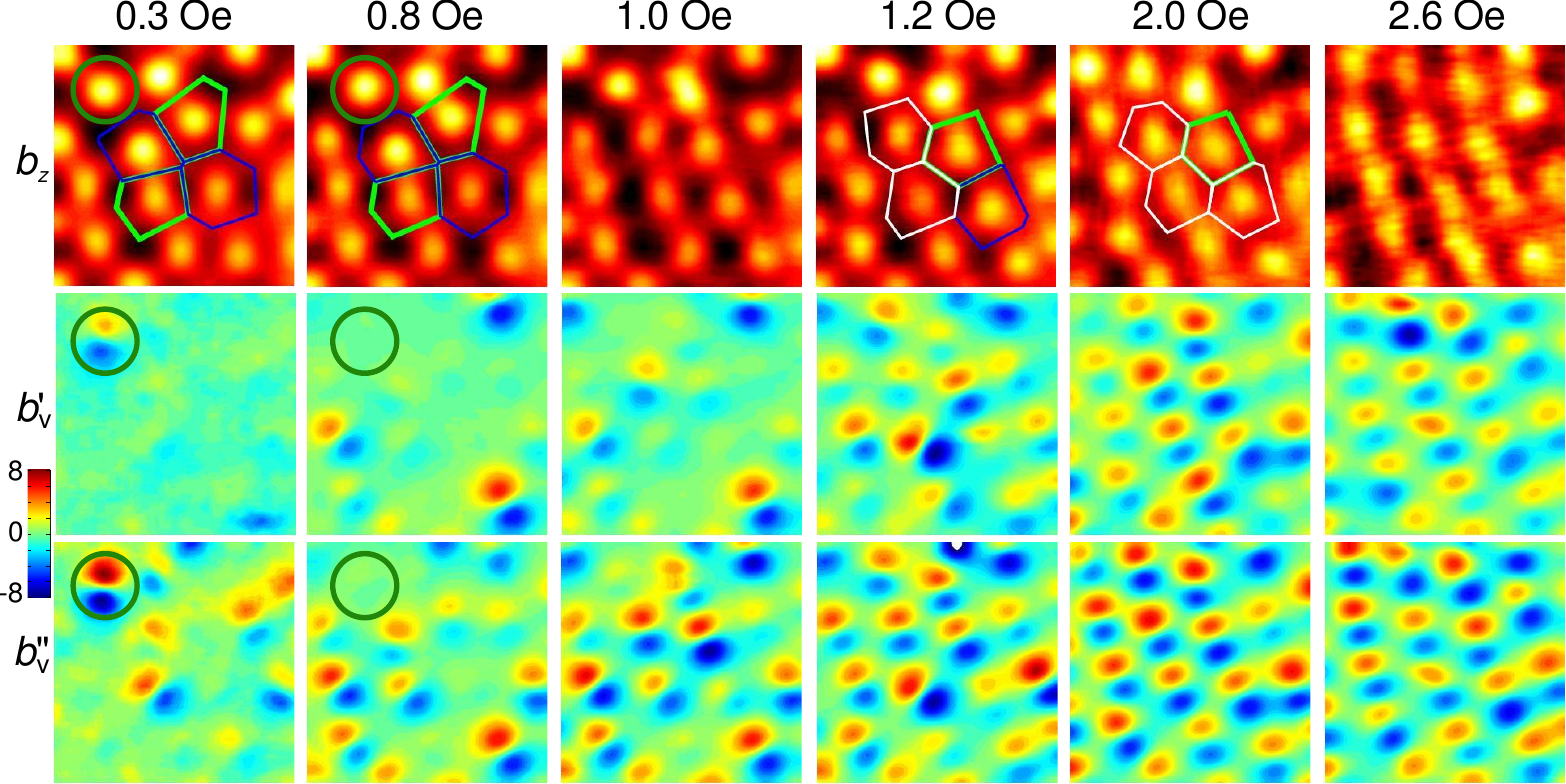}
\caption{(Top row) Scanning Hall probe microscopy images of the local induction, $b_z(x,y)$, acquired at a temperature of $T=6.8$ K and a dc magnetic field $H=1.0$ Oe while shaking with an external applied ac field of frequency $f=77.123$ Hz and increasing amplitude (left to right) $H_{ac}=0.3$, 0.8, 1.0, 1.2, and 2.6 Oe. The polygons are Voronoi constructions indicating the coordination number $\nu$ of each of the four central vortices (green for $\nu=5$, white for $\nu=6$ and blue for $\nu=7$). (Center and bottom rows) Simultaneously acquired maps of (center) in-phase, $b_v'(x,y)$, and (bottom) out-of-phase, $b_v''(x,y)$, response components. To unify the color map scale, we normalized both $b_v'$ and $b_v''$ by $10^{-3} H_{ac}$ (in gaussian units).
} \label{fig.ac-exp-images}
\end{figure*}
Figure \ref{fig.ac-exp-images} summarizes the evolution of the initial disordered vortex state upon increasing $H_{ac}$ at the same experimental conditions of Figure \ref{fig.ExpMean}. From left to right, the top row in Fig.\ref{fig.ac-exp-images} shows a representative subset of acquired time-averaged vortex distributions for  $H_{ac}=0.3$, 0.8, 1.0, 1.2, 2.0 and 2.6 Oe, respectively. These images are obtained by measuring the time-averaged local magnetic induction, $\langle b_z(x,y,t)\rangle_t$ with SHPM, while the vortices react to the applied ac magnetic field. As it is clear, the scan area comprises about 16 vortices which never leave the image within the investigated amplitude range, thus justifying the application of the procedure adopted to remove the background. The middle and bottom panels show the simultaneously acquired in-phase, $b_v'(x,y)$, and out-of-phase, $b_v''(x,y)$, response components.

The evolution of the vortex pattern as well as the motion of individual vortices reveal a picture far richer than that suggested by the simple linear behaviour of the background response. For small amplitudes, $H_{ac}<1.0$ Oe, the average vortex positions remain unaltered from the original disordered FC state. In spite of that, the local ac response indicates that more and more vortices participate on the dynamics as $H_{ac}$ is increased from 0.3 to 0.8 Oe. Moreover, in this amplitude range, vortices shake in different directions and with different amplitudes. This is in strong contrast with mean field models of vortex response, which assume vortices to shake in the same direction that the applied Lorentz force. This finding, can be understood as a result of the local anisotropy of vortex-vortex interactions and the disorder of the restoring force strength probed by each vortex. As a startling result, some vortices seem to simply stop moving at a higher excitation (Encircled vortex in Fig.\ref{fig.ac-exp-images}, $H_{ac}=0.3$ and $0.8$ Oe) suggesting that tiny changes in vortex positions can change considerably the energy landscape probed by the vortex, either because of different pinning
conditions or a different excited mode of the vortex array.  %In

At $H_{ac}=1.0$ Oe, the vortex configuration changes dramatically to
a more ordered state.
As the amplitude is further increased, the vortex arrangement progressively acquires the triangular symmetry. Concomitantly, the shaking directions become more correlated to each other and parallel to a principal axis of the triangular lattice. Notice that up to $H_{ac}=2.0$ Oe the maxima of magnetic induction are well-defined indicating that the shaking amplitude of all vortices is much smaller than the lattice constant. At $H_{ac}=2.6$ Oe, the dc image is considerably blurred along a particular direction. However, for this amplitude range the ac-images present a highly correlated motion. This indicates that the travel range of a single vortex does not exceed the lattice constant, but is sufficiently large to result in a low and blurred time averaged vortex signal. A similar transition in the dynamical properties of the same prepared state was observed upon increasing the temperature while keeping $H$ and $H_{ac}$ constant.

To better quantify the amount of order in the local vortex configuration, we have calculated the coordination number of each of the four central vortices in the image. The calculation consists of finding all local maxima in the images, which we identify as the mean vortex position, and then performing a Voronoi construction. For $H_{ac}<1.0$ Oe, all four central vortices can be identified as either 5-fold or 7-fold disclinations comprising part of a probably larger cluster of topological defects. Above 1.0 Oe, these defects are gradually healed, becoming sixfold coordinated vortices as the vortex arrangement approaches a triangular lattice. At 2.6 Oe, the image is considerably blurred by the shaking of vortices in a way that the Voronoi construction cannot be performed accurately.

It is worth mentioning that within the whole amplitude range the imaginary component of the vortex response is considerably larger than the real component, indicating that thermally assisted vortex hopping plays a major role during the ac shaking. However, since the response is highly nonlinear in a broad amplitude range, applying any of the known mean-field models to extract the hopping time is hardly justifiable.

\section{Numerical simulations}
\label{sec.MD}

\subsection{Model and numerical details}

In order to gain additional insight over the mechanisms behind the experimentally observed strong dissipation, dynamical reordering and self-organization, we performed molecular dynamics simulations of vortices interacting with a random pinning potential in a 2D superconducting system. The vortex-vortex pair potential is modelled as $U_{vv}({r}_{ij})=\epsilon K_0(r_{ij}/\lambda)$, where $\epsilon=\phi_0^2/(2\pi\mu_0\lambda)$ is the energy scale. Here, we take $\lambda(0)=150$ nm (typical for NbSe$_2$) and a reduced temperature $T/T_c=0.965= 6.8$ K$/7.05$ K, which sets our length scale to $\lambda=0.80$ $\mu$m. The magnetic flux density is fixed at $B=1.0$ Oe, which is equivalent to a vortex density $n_v=0.037$ $\lambda^{-2}$ similar to the experiment.  The disorder induced by pinning sites is modelled by a Gaussian-correlated, random landscape of root-mean-square (rms) value $U_0$ and a correlation length (typical inter-valley length scale) $\xi_p$. Because vortices can not resolve distances smaller than the coherence length we choose $\xi_p=0.06625\lambda$, which is close to the typical value of $\xi$ for NbSe$_2$ at a reduced temperature $T/T_c = 0.965$. The resulting pinning potential, $U_p(x,y)$, represents the superposition of a high density ($n_p\gtrsim\xi_p^{-2}=228\lambda^{-2}$) of randomly distributed point defects. In contrast to previous models~\cite{Valenzuela2002, Mangan2008, Moretti2004, Chandran2004}, where a diluted distribution of pinning centers was used, here individual pinning potentials do overlap considerably, which we believe represent a more realistic scenario for weak-pinning materials like NbSe$_2$.

The dynamics of the vortex system subjected to an ac drive ${\mathbf F}(t)={\mathbf A}\cos\omega t$ is simulated by a standard Langevin dynamics algorithm, which essentially corresponds to numerically integrating the overdamped equation of motion:
\begin{equation}\label{eq.motion}
\eta\dot{\mathbf{r}}_i(t) = - \nabla_i E_p - \nabla_i E_{vv} +\mathbf{F}(t)+\mathbf{\Gamma}_i(t) ,
\end{equation}
where $\mathbf{r}_i(t)$ is the vortex position and $\nabla_i$ the gradient operator with respect to $\mathbf{r}_i(t)$. Here $E_p=\sum_jU_p(\mathbf{r}_j)$ is the total pinning energy, $E_{vv}=\frac{1}{2}\sum_{jk} U_{vv}({r}_{jk})$ is the total vortex-vortex interaction energy and $\mathbf{\Gamma}_i(t)$ is the Langevin force, representing thermal fluctuations of the vortices. A possible inertial term, not shown in Eq.~(\ref{eq.motion}), is accepted to be very small so that there is a short (negligible) initial period of acceleration needed to reach the steady state motion we consider.

 We ran simulations on a rectangular box of size $L_x\times L_y$ ($L_y=\sqrt{3}L_x/2$) with periodic boundary conditions for system sizes $L_x=60\lambda$ (48 $\mu$m), $120\lambda$ (96 $\mu$m) and $180\lambda$ (144 $\mu$m). All results discussed below are qualitatively the same for all investigated system sizes. Therefore, we will present only results from the smaller system ($60\times51.96$ $\lambda^2$) for which a more detailed analysis was performed.

Before analyzing the ac vortex response, we thermalize the vortex distribution following a simulated annealing scheme, where the Langevin force in Eq.~(\ref{eq.motion}) is slowly decreased down to zero. This way, vortices are settled in a low energy configuration. The value of $U_0$ was chosen in a way as to result in a weakly disordered vortex lattice even at zero temperature. A typical configuration is shown in Fig.~\ref{fig:VLconf}(d). In what follows we take $U_0=1.9\times10^{-4}\epsilon$, which, as shown below, results in a vortex configuration with $17.4$\% of defects. The corresponding pinning coupling constant is estimated as $\alpha_p=6.03\times10^{-2}\,\epsilon/\lambda^2$. The calculated value of $U_0=0.8$meV and $\alpha_p=6.33\times10^{-8}$N$/$m are at least one order of magnitude smaller than what is found in literature for strong pinning Nb films\cite{Park1992} at 4.5 K and Pb films\cite{Raes2012} at 6.9 K. Subsequently, the temperature is fixed at the desired value and a uniform ac excitation of angular frequency $\omega$ is applied. The response of the vortex system is then studied as a function of excitation amplitude. In all calculations we used a drive period $P=10^5t_0$, corresponding to a frequency $\omega=6.28\times 10^{-5}\,t_0^{-1}$ much smaller than the typical pinning frequency $\omega_p=\alpha_p/\eta=6.03\times10^{-2}\,t_0^{-1}$. We run simulations for different drive orientations $\theta$ and observed that the results are qualitatively independent on this parameter. All results presented below correspond to $\theta=-60^\circ$ with respect to the horizontal axis. The initial configurations correspond to a temperature below the freezing point $T_f\simeq 3.3U_0$, below which the vortex configuration keeps unchanged at zero drive.

In order to characterize the response of each individual vortex to the ac shaking we compute the in-phase, $\zeta_{i1}'$, and out-of-phase, $\zeta_{i1}''$, components of its displacement in direction $\zeta=$ $x$ or $y$:
\begin{equation}\label{vortexdisplacements}
\zeta_{i1}' + i\zeta_{i1}''=\frac{2}{\Delta t}\!\int_0^{\Delta t}\! dt\,\zeta(t)[\cos\omega t + i\sin\omega t]
\end{equation}
where $\Delta t$ is the measuring time corresponding to an integer number of periods. Here we took $\Delta t=5P$. The overall response is quantified by the mean in-phase and out-of-phase components of vortex displacements, defined as
\begin{equation}
u_1=u'_1+iu''_1=\frac{1}{N}\!\sum_{j=1}^N\left[\sqrt{x'^2_{j1} + y'^2_{j1}} + i\sqrt{x''^2_{j1} + y''^2_{j1}}\right]
\end{equation}
The topological order is evaluated by counting the number of topological defects (vortices with coordination number different from 6 and averaging it over the measuring time $\Delta t$). The orientational order is characterized by the real part of the sixfold bond-angle order parameter averaged over all vortices and time, $\psi_6=\frac{1}{\Delta t}\!\int_0^{\Delta t}\! dt \,\frac{1}{N}\!\sum_k\sum_{j\in {\rm Neigh}(k)}e^{6i\theta_{kj}}$. The time integrals were performed after an interval of typically 10 periods.

\subsection{Dynamical reordering}

\begin{figure*}[t]
\hspace{2.54cm}\includegraphics[width=0.6\linewidth]{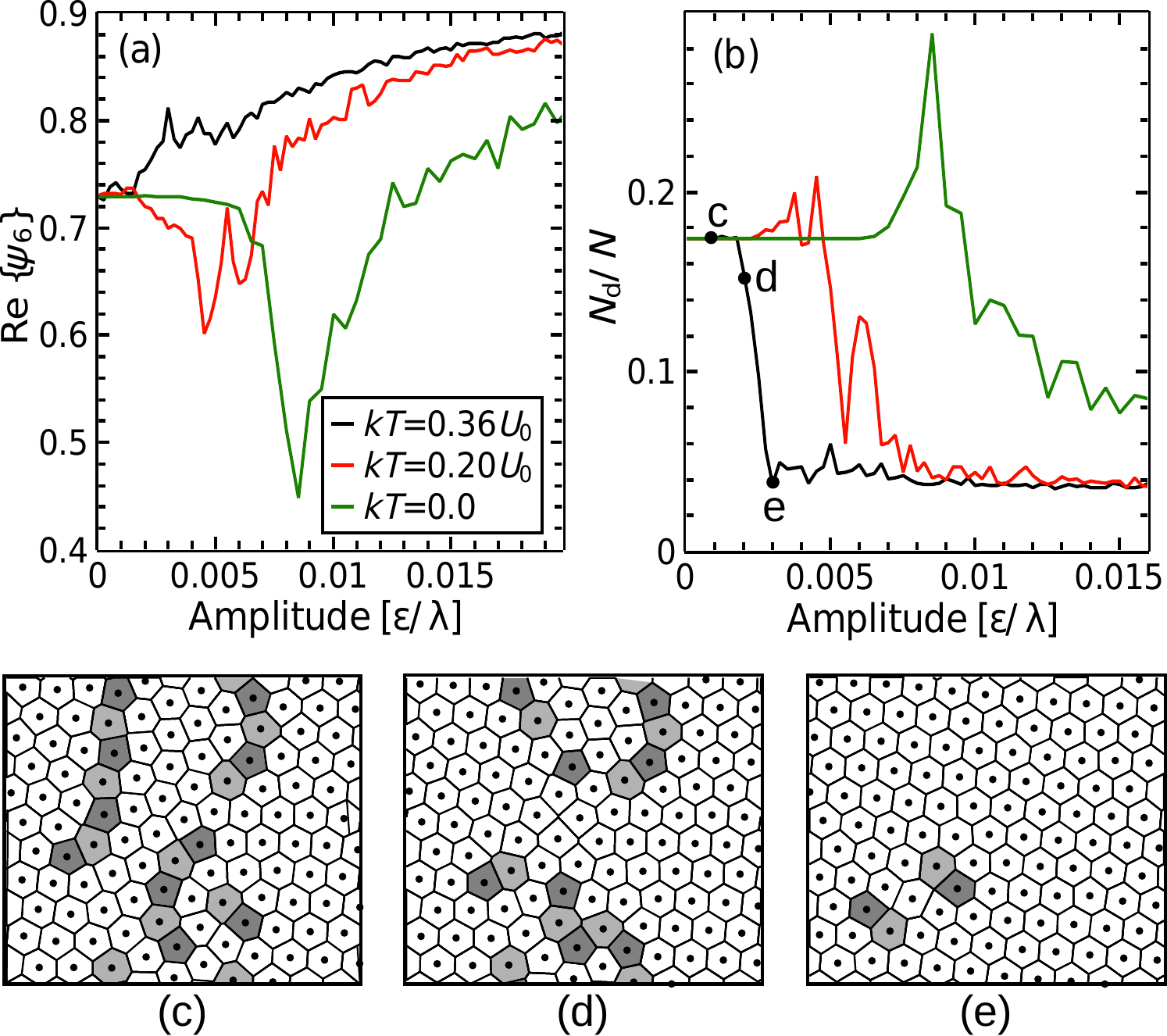} \vspace{-5mm} \caption{(a), (b): Amplitude dependence of (a) the real part of the sixfold bond-angle order parameter averaged over all vortices and time and (b) time-averaged number of defects normalized by the total number of vortices. The curves are plotted for three different temperature values as labeled in panel (a). (c-e) Snapshots of vortex configurations and Voronoi constructions for the points indicated in panel (b). Defects of positive (negative) topological charge are depicted in light (dark) gray.} \label{fig:MDreorder}
\end{figure*}

\begin{figure*}[t]
\hspace{2.54cm}\includegraphics[width=0.65\linewidth]{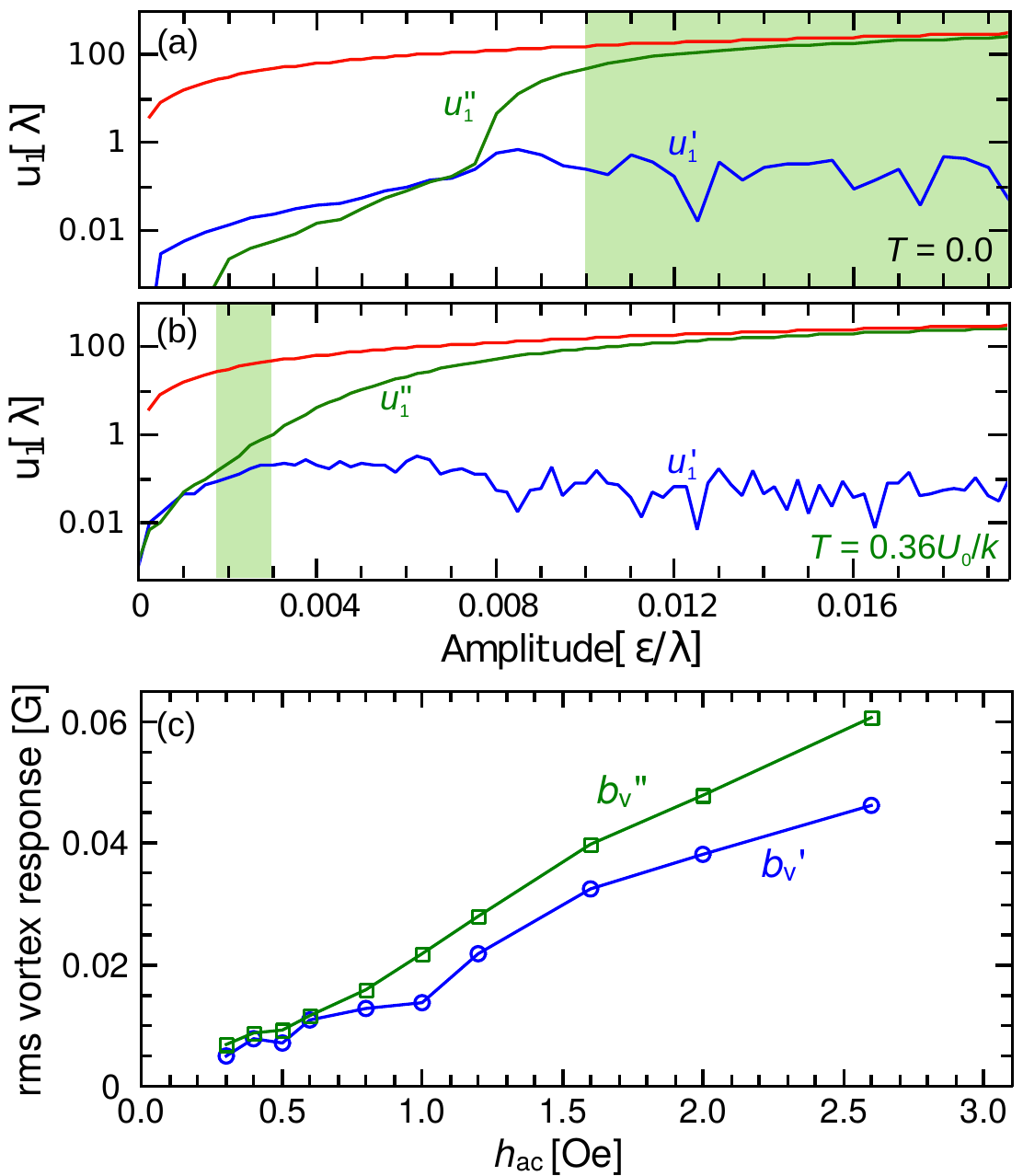}
\caption{(a and b) Real ($u_1'$) and imaginary ($u_1''$) parts of the first harmonic of the vortex displacements averaged over all vortices for $T=0$ (a) and $T=0.36U_0/k$(b). The shaded regions correspond to the amplitude range where the defect density effectively decrease below its static value reaching a minimum value. For comparison, we show the response predicted for the flux-flow regime, $u=iA/\eta\omega$ (red dashed curve).  (c) Root-mean-square values of the in-phase (circles) and out-of-phase (squares) components of vortex response within the scan area as functions of $h_{ac}$.
} \label{fig.u1-Fk}
\end{figure*}

Fig.~\ref{fig:MDreorder}
shows for different temperatures the evolution of orientational order (quantified by $\Re{\{\psi_6\}}$) and number of defects ($N_d$) as the drive amplitude is increased [panels (a) and (b), respectively]. For $T=0$, the healing of defects is preceded by an increase in $N_d$ and concomitant deterioration of the orientational order, which is a signature of plastic dynamics, with strong relative motion between vortices \cite{Duarte1996,Ryu1996,Olson1998}. Only at an amplitude somewhat larger, $A\simeq0.01$, the number of defects decreases below its zero amplitude value. At this amplitude and above vortices move over several vortex-lattice spacings, as revealed by the in-phase and out-of-phase vortex displacements, $u'_1$ and $u''_1$, shown in figure~\ref{fig.u1-Fk}(a),
($|u_1|\gtrsim 10^2 \lambda \sim 18$ lattice spacings). It then becomes clear that, for $T=0$ and low drive frequency, the mechanism behind the healing of defects is similar to the plastic-to-elastic transition of vortices moving under a dc drive and thereby related to the dynamical washout of the pinning potential. Indeed, the response in this amplitude range approaches that of the pin-free flux-flow regime ($u=iA/\eta\omega$).

A very different picture emerges for $T>0$. Here the appropriate time scale is roughly given by $\tau_r=\tau_pe^{U_0/kT}$, which depends upon the ratio $U_0/kT$ and leads to a characteristic frequency $\omega_r=\tau_r^{-1}\ll\omega_p$, comparable to the low frequencies considered here. For $T=0.36U_0/k$, $\omega_r/\omega\approx 0.21$. Clearly, the hopping dynamics, responsible for the relaxation mechanism, plays an important role in the response. Indeed, for such temperature, at the onset of reordering, the healing of defects is much faster than for $T=0$ and there is no proliferation of defects preceding the ordering transition. This means that, upon increasing amplitude, the vortex array goes through a quick transition from a pinned disordered lattice to a moving elastic phase, with no intermediate plastic phase. In contrast to the $T=0$ case, here, for amplitudes close to the ordering transition, vortex excursions are restricted to distances considerably smaller than $\lambda$ [see behavior of $u_1$ in  Fig.~\ref{fig.u1-Fk}(b)], pointing to an entirely different ordering mechanism, ruled by thermal hopping of vortices.

In Fig.~\ref{fig.u1-Fk}(c) we plot the rms values of the in-phase and out-of-phase components of the vortex response as a function of amplitude as derived from Fig.\ref{fig.ac-exp-images} within a single scan area.  A very good qualitative agreement is observed with the vortex response predicted by MD simulations in the green shaded area in Fig.~\ref{fig.u1-Fk}(a), which corresponds with the amplitude regime where reordering takes place and where the ac-experiments are performed. As such, it confirms the MD model is able to capture the main physical ingredients ruling the vortex dynamics in our system.

\subsection{Trajectories of a single vortex}

In previous works \cite{Kramer,Kramer1,Raes2012}, we have demonstrated that the SSM technique is capable of extracting useful information of individual vortex dynamics  with single vortex resolution. Here we will benefit from the MD simulation to analyse the vortex trajectories at much smaller scales. Let us first establish a connection between our theoretical calculations and the experimental observations on a local scale. To that end, we calculated the first harmonic of the local flux density, $b_z({\mathbf r},t)$, induced by the ac vortex dynamics at a distance $z_0=0.5\lambda$ away from the sample surface:
\begin{equation}
b'_v(x,y) + ib''_v(x,y)=\!\frac{2}{\Delta t}\!\int_0^{\Delta t}\hspace{-4mm}dt\,b({\mathbf r},t)[\cos\omega t + i\sin\omega t]
\end{equation}

\begin{figure*}[t]
\hspace{2.54cm}\includegraphics[width=0.75\linewidth]{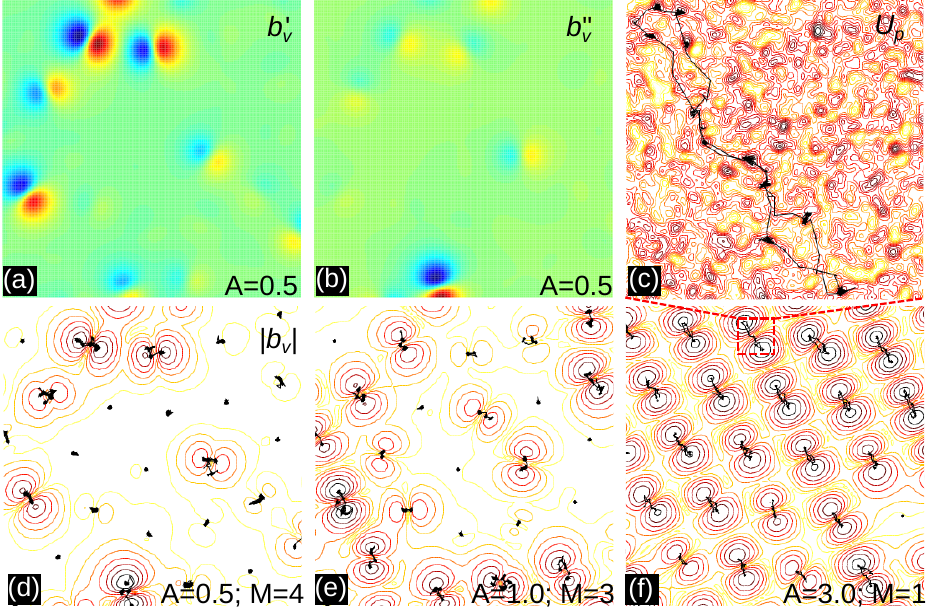} \caption{(a) and (b): density plots of the real (a) and imaginary (b) components of the first harmonic of the vortex flux distribution, $b_v$, in a 26$\times$26 $\lambda^2$ area of the sample for a driving force amplitude $A=0.5\times10^{-3}\epsilon/\lambda$. (d)-(f): contour plots of the absolute value of $b_v$ for drive amplitudes $A=0.5$, 1.0 and 3.0 (in units of $\times10^{-3}\epsilon/\lambda$) respectively. Individual vortex trajectories during one forcing period are also shown. For better visualization, all vortex trajectories with respect to their mean position were magnified by a factor $M=4$ (d) and $M=3$ (e). Panel (c) is a zoom-in of the 2.6$\times$2.6 $\lambda^2$ region depicted in (f) and shows one vortex trajectory and the pinning landscape nearby.} \label{fig:MDimages}
\end{figure*}

These results are shown in Fig.~\ref{fig:MDimages}. The contribution of each vortex to the flux density at the probe position ${\mathbf r}=(x,y,z_0)$ and instant $t$ was accounted for by using the monopole approximation for a vortex flux profile~\cite{CarneiroBrandt2002}. The results are presented for $T=0.36U_0/k$ and a few amplitude values. Notice the remarkable resemblance with the experimental data shown in Fig.~\ref{fig.ac-exp-images}.

For comparison, we also plot the vortex trajectories, which reveal that, despite the rather erratic dynamics of the vortices, specially at smaller drive amplitudes, their main direction of motion can be captured by the absolute value of $b_v(x,y)$.  Moreover, the intensity of peaks and valleys of $b_v$ near a given vortex is, in general, proportional to the amplitude of motion of that vortex (c.f. the Appendix).  This allows us to witness the pronounced uncorrelated dynamics at small drive amplitude [panels (a), (b), (d) and (e)], where vortices shake with very different amplitudes and directions (away from the drive direction), in excellent agreement with the experimental data. In contrast, when reordering takes place [panel (f)] the motion becomes more uniform (i.e. higher correlation) and both the direction of motion and the main axis of the triangular lattice align themselves to the drive direction, also in agreement with our experimental observations.

A closer look at a typical vortex trajectory is presented in panel (c) together with a contour plot of the local pinning landscape. The trajectories were plotted by tracing the position of the vortex every $10^3$ time steps. This makes it possible to observe that vortices spend much more time trapped by some favorable pinning centers than traveling between them. Therefore the dynamics under those conditions is essentially ruled by hoping of vortices between the
most favorable pinning sites. %It is worth noting that
%this long trapping time is consistent with the observed negligible
%difference between kinetic and static friction forces for this amplitude range.
These observations are in agreement with recent scanning tunneling microscopy experiments on similar NbSe$_2$ crystals having a much denser flux line lattice\cite{Timmermans2014}.

\section{Conclusions}

In conclusion, in this work we investigated the local ac dynamics of a disordered vortex state upon increasing drive  by a combination of two local probing techniques, scanning Hall probe microscopy and scanning susceptibility microscopy. Our experimental data provided direct evidence of dynamical healing of topological defects as the ac excitation amplitude is increased. Moreover, the SSM images revealed two very different behaviors of the individual vortex response: uncorrelated dynamics, where vortices shake at different directions with different amplitudes, and correlated dynamics, where, upon the healing of defects, the directions of motion of all vortices align and they respond almost in unison. The observed microscopic dynamics is confronted to the extensively used phenomenological microscopic models of vortex motion proposed to explain the macroscopic response. We show that the approximations made in these models represent an oversimplification of a much richer ac dynamics.  Molecular dynamics simulations are used to gain further insight in the thermally driven organisation of the vortex motion and allow us to visualize the vortex trajectories otherwise hidden by the limited resolution of the local probe techniques.

% Specify following sections are appendices. Use \appendix* if there
% only one appendix.
\appendix*

\section{Visualization of the vortex dynamics using the SSM technique}

Here we show explicitly that the main direction of motion can be captured by the absolute value of $b_v(x,y)$, measured in the SSM experiments. We denote with ${b_z}^v({\mathbf r},t)$ the magnetic induction carried by a single vortex, shaking back and forth around its equilibrium position, ${\mathbf r}_{i0}$. When the deviations from equilibrium are small, we can expand the flux density carried by a single vortex around ${\mathbf r}_i={\mathbf r}_{i0}$
\begin{equation}\label{EqA1}
b_z({\mathbf r},t)=b_z({\mathbf r}-{\mathbf r}_i)=\sum_{p=0}^\infty(-\delta{\mathbf r}_i \cdot \nabla)^p \frac{ b_z({\mathbf r}-{\mathbf r}_{i0})}{p!},
\end{equation}
where $\delta{\mathbf r}_i={\mathbf r}_i-{\mathbf r}_{i0}$ is the vortex displacement. By keeping terms up to the second order in the vortex displacements, it can be shown that the first Fourier component of the vortex flux density is given by,
\begin{equation}\label{EqA2}
b_v(x,y) = -\delta{\mathbf r}_{1i} \cdot \nabla b_z({\mathbf r}-{\mathbf r}_{i0}).
\end{equation}
where, $\delta{\mathbf r}_{1i}=(x_{1i}, y_{1i})$  is the complex vortex displacement as defined in Eq.~\ref{vortexdisplacements}. In other words, within second order approximation, \emph{$b_v$ is just the directional derivative of the flux induced by the vortex at its equilibrium position}. Therefore, the direction of strong gradients in $b_v$ can be identified as the direction of the vortex response (see Eq.~\ref{EqA2}). Notice that the length scale in the case of a diluted vortex distribution for $\nabla b_z({\mathbf r}-{\mathbf r}_{i0})$ is the penetration depth. This scale exceeds, in the linear regime, typical vortex displacements and hence one can safely keep the leading order terms in Eq.~\ref{EqA1}.

% If you have acknowledgments, this puts in the proper section head.
\section{Acknowledgements}

This work was partially supported by the Methusalem Funding of the Flemish Government, the Fund for Scientific Research-Flanders (FWO-Vlaanderen), the Fonds de la Recherche Scientifique - FNRS and the Brazilian funding agencies CNPq and FACEPE, and the program for scientific cooperation FRS-FNRS-CNPq. The work of A.V.S. is partially supported by ``Mandat d'Impulsion Scientifique" of the F.R.S.-FNRS and the by "Cr\'edit de d\'emarrage", U.Lg. The Authors would like to thank J. Ge for providing us with the NbSe$_2$ crystal.

% Create the reference section using BibTeX:
\bibliography{arxiv_references}

\end{document}